\documentclass[twocolumn,floatfix,preprintnumbers,superscriptaddress]{revtex4}

\usepackage[utf8]{inputenc}
\usepackage[colorlinks=true,citecolor=blue,linkcolor=blue]{hyperref}
\usepackage[normalem]{ulem}
\usepackage{url}
\usepackage{graphicx,wrapfig,float,slashed,cancel}
\usepackage{amsmath,amssymb,epsfig,graphicx,xcolor,stmaryrd}
\usepackage{bm}
\usepackage{enumitem}
\usepackage{xcolor}

\definecolor{darkblue}{RGB}{1, 90, 173}

\begin{document}


\title{Spectroscopic parameters and electromagnetic form factor of kaon in vacuum and a dense medium}

\author{N. Er\textsuperscript}
\email{nuray@ibu.edu.tr}
\affiliation{Department of Physics, Bolu  Abant \.{I}zzet Baysal University,
G\"olk\"oy Kamp\"us\"u, 14030 Bolu, Turkey}

\author{K. Azizi} %
\email{kazem.azizi@ut.ac.ir(Corresponding Author)}
\affiliation{Department of Physics, University of Tehran, North Karegar Ave. Tehran 14395-547, Iran}
\affiliation{Department of Physics, Do\v{g}u\c{s} University, Dudullu-\"{U}mraniye, 34775
Istanbul, Turkey}

\date{\today}

\preprint{}

\begin{abstract}
	The spectroscopic parameters as well as electromagnetic form factor of the strange particle kaon are investigated in vacuum and  a medium with finite density. The obtained vacuum mass and decay constant, which are consistent with the existing experimental results, are used to extract the $ Q^2 $ dependence of the kaon electromagnetic form factor in the interval $Q^2\in [0,10]$ GeV$^2$ in vacuum. The obtained results at lower and intermediate values of  $ Q^2 $ are consistent with the existing experimental data within the presented uncertainties. The $ Q^2 $ behavior of the electromagnetic form factor of kaon in vacuum and in the interval $Q^2\in [0,6]$ GeV$^2$ is in a nice agreement with the existing predictions of the Lattice QCD and the solution of the Bethe-Salpeter equation for the model of Nambu and Jona-Lasinio (NJL) with proper-time regularization, as well. The obtained vacuum radius for kaon is also in a nice agreement with the world's average experimental result. We extend the analyses to a medium with higher densities and obtain the behavior of the mass, decay constant, electromagnetic form factor and radius with respect to density. The obtained results for some of the parameters are compared with the existing predictions of other models and approaches. The results obtained in the present study can be useful for future experimental and theoretical studies both in vacuum and a dense medium.
\end{abstract}


\maketitle

\section{Introduction} \label{sec:intro}

Investigation of the light pseudoscalar mesons with spin-parity $J^P=0^-$  plays an important role  for a better understanding of the perturbative and nonperturbative natures of Quantum Chromodynamics (QCD). The spectroscopic properties of these particles have been widely studied theoretically and experimentally in vacuum. Although, there have been many efforts to  study the electromagnetic form factors, inner charge and magnetization distributions and mechanical properties of these particles such as their radius, we still need more experimental and theoretical studies to clarify the obtained results in vacuum. Theoretical calculations of the above mentioned parameters in a dense medium can be very useful for analyses of the future data provided by the in-medium experiments.

In terms of experimental studies, the simplest particles among light mesons are the pion triplet. In the last years, the static properties and dynamical features of pions have been extensively studied in many works \cite{PhysRevLett.43.246,PhysRevLett.61.2296,REINHARD,Blin,PhysRevC.86.038202,Bakker,Melikhov,Leitner,Brodsky,Nguyen,Dominguez,Dorokhov,Truedsson,deMelo:2014gea}. The next simple meson available for experimental studies is the strange particle, kaon,  but  there are relatively fewer studies devoted to the study of parameters of kaons \cite{Mishra2010,Maris:2000sk,PhysRevD.96.034024,PhysRevC.81.035203,Krutov:2016luz,Solomey,Carrasco:2014poa,Dally:1980dj,Amendolia:1986ui,Yabusaki:2017sgs}.

The electromagnetic properties such as electromagnetic form factors (FFs) of particles carry useful information about the distributions of the  charge and magnetization, determination of which  can help us obtain valuable knowledge on the internal structure of hadrons in terms of quark-gluon organizations and their geometric shapes.  Theoretical calculations as well as experimental studies on the electromagnetic properties of nucleons and light mesons are done by different collaborations and experimental groups. Different groups consider different ranges for the transferred momentum squared $ Q^2 $ carried by the electromagnetic current. We need to extract the FFs in a wide range of   $ Q^2 $ to get a complete knowledge on the FFs of hadrons. In the present study, we determine the FFs of the strange meson $ K^+ $ in the interval $Q^2\in [0,10]~GeV^2$ using the QCD sum rules in vacuum. The mass and decay constant of the kaon are entered as the main input parameters  to the calculations. We use vacuum two-point sum rules to extract these spectroscopic parameters. We discuss the behavior of the kaon electromagnetic form factor with respect to  $ Q^2 $  and compare the obtained results with the existing experimental data and predictions of other theoretical and phenomenological models. We calculate the vacuum charge radius of $ K^+ $ particle and compare the result with the experimental data and existing theoretical predictions.

Investigation of hadronic parameters at a dense medium, on the other hand, is very important to complete our knowledge on the nature and structure of hadrons both theoretically and experimentally. By the progresses made in the experimental side, we hope it will be possible to study the statistic and dynamical properties of hadrons in future in-medium experiments. Motivated by this, in the present study, we extend our calculations on the spectroscopic parameters as well as electromagnetic FFs and charge radius of $ K^+ $ to a dense medium and discuss the variations of these hadronic parameters with respect to the density of the medium. To obtain our results we use the two-point in-medium sum rules for the spectroscopic parameters and three-point in-medium sum rules for the in-medium electromagnetic interaction to extract the electromagnetic FFs.  There are some limited information on some of these parameters available in the literature that we compare our results with these information. 

The paper is organized in the following way. In next section, we calculate the mass and decay constant of $ K^+ $ both in vacuum and a dense medium and obtaine their numerical values. In section III, we calculate the electromagnetic form factor and charge radius of $ K^+ $  again in vacuum and a dense medium and perform our numerical analyses. Last section is devoted to the summary and our concluding notes.

\section{The spectroscopic parameters of Kaon in vacuum and a dense medium}

The in-medium time-ordered two point correlation amplitude $\Pi_{\mu\nu}$, responsible  for the calculation of  mass and decay constant of kaon, is given as 
\begin{equation}
\label{masss}
\Pi_{\mu\nu} = i \int d^4 x e^{ip\cdot x} \langle\psi_0|\mathcal{T}[j_{\mu}(x)j^{\dagger}_{\nu}(0)]|\psi_0\rangle,
\end{equation}
where $|\psi_0\rangle$ is the ground state of the nuclear medium.  For the vacuum calculations, this state is replaced with the vacuum one  $|0\rangle$. The interpolating current of kaon in Eq. (\ref{masss}) is
\begin{equation}
\label{current}
j_{\mu}(x)=\bar{s}^a(x)\gamma_{\mu}\gamma_{5} u^a(x),
\end{equation}
and $\mathcal{T}$ denotes the time ordered product of two currents.

In accordance with the  QCD sum rule method, the correlation function in  Eq. (\ref{masss})  can  be represented in terms of hadronic  parameters called the physical or hadronic side, and alternatively in terms of the QCD degrees of freedom called the QCD side. The hadronic side is obtained in terms of the the in-medium decay constant $f_K^*$ and mass $m_K^*$  of kaon. The axial current considered for the kaon simultaneously couples to both the pseudoscalar ($ PS $)  and axial vector ($ AV $)  kaons. Hence, after performing the four space-time integral and isolating these two states, we get
\begin{eqnarray}
\label{corre3}
\Pi^{Phys}_{\mu\nu} &=&\frac{ \langle\psi_0|j_{\mu}|K_{AV}(p^*)\rangle  \langle K_{AV} (p^*)|j^{\dagger}_{\nu}||\psi_0\rangle}{m_{AV}^{*2}-p^{*2}}\nonumber\\
&+&\frac{ \langle\psi_0|j_{\mu}|K_{PS}(p^*)\rangle  \langle K_{PS} (p^*)|j^{\dagger}_{\nu}||\psi_0\rangle}{m_{PS}^{*2}-p^{*2}} + ...,
\end{eqnarray}
where dots stand for contributions of the higher resonances and continuum states in both the $ PS $ and $ AV $ channels. 
The in-medium momentum is given as $ p^*_{\mu}=p_{\mu}-\Sigma_{\upsilon }u_{\mu}$,
where $ \Sigma_{\upsilon } $ is the vector self-energy and  $u_{\mu}$ is the four-velocity of the nuclear medium. We shall work in the rest frame of the medium, $u_{\mu}=(1,0)$.

To simplify the above relation, we define the corresponding matrix elements in terms of the in-medium masses, decay constants, momenta and polarization vector of $ AV $ state: 
\begin{eqnarray}
\label{defff}
 \langle\psi_0|j_{\mu}|K_{AV}(p^*)\rangle &=& f^*_{AV} m^*_{AV} \epsilon^*_{\mu}, \nonumber \\
 \langle\psi_0|j_{\mu}|K_{PS}(p^*)\rangle &=& f^*_{PS} m^*_{PS} p^*_{\mu}.
\end{eqnarray}
In vacuum, the in-medium masses, decay constants and momenta as well as the polarization vector of axial kaon are replaced by their vacuum versions without stars. Using Eq. (\ref{defff}) in Eq. (\ref{corre3}) leads to
\begin{eqnarray}
\label{correfinal}
\Pi^{Phys}_{\mu\nu} &=& \frac{ f_{AV}^{*2} m_{AV}^{*2}} { m_{AV}^{*2}-p^{*2}}\Big[-g_{\mu\nu}+\frac{p_{\mu}^{*}p_{\nu}^{*}}{p^{*2}}\Big] \nonumber \\
&+&  \frac{ f_{PS}^{*2} m_{PS}^{*2}p_{\mu}^{*}p_{\nu}^{*}} { m_{PS}^{*2}-p^{*2}} + \cdots.
\end{eqnarray}
We need to extract the contribution of  $ PS $ state only. 
To remove axial particle contributions in Eq. (\ref{correfinal}) we multiply the  physical side by $\frac{p_{\mu}p_{\nu}}{p^2}$ for vacuum and  $p^*_{\mu}$ for in-medium cases.  In the dense medium, we need more structures and some rules to find some parameters beyond the vacuum: like the vector self energy. Constructing more sum rules allows us to simultaneously find the mass, decay constant and vector self energy of the particle.  The Borel transformation with respect to momentum squared is applied to suppress the contributions of the higher states and continuum. After removing the contributions of $ AV $  states and performing the standard calculations,   the Borel transformed physical side for vacuum and dense medium are obtained as
\begin{eqnarray}
\label{corre5}
\mathbf{ \hat{B}}\Big[\frac{p_{\mu}p_{\nu}}{p^2}\Pi^{Phys}_{\mu\nu}\Big] &=& f_{PS}^{2}m_{PS}^{4} e^{-m_{PS}^2/M^2}  \mathbf{I}, 
\end{eqnarray}
and 
\begin{eqnarray}
\mathbf{ \hat{B}}\big[p^*_{\mu}\Pi^{Phys}_{\mu\nu}\big] &=& f_{PS}^{*2}m_{PS}^{*4} e^{-\mu^2/M^2} (p_{\nu}-\Sigma_{\upsilon} u_{\nu}),\nonumber\\
\end{eqnarray}
where $\mu^2=m_{PS}^{*2}-\Sigma^2_{\upsilon} +2 \Sigma_{\upsilon} p_0$, $M^2$ is the Borel mass parameter, $ p_0 $ is the energy of the quasi-particle in medium and $ \mathbf{I} $ denotes the unit matrix.

On the other hand, the correlation function in Eq. (\ref{masss}) can be calculated in terms of quarks and gluon degrees of freedom in deep Euclidean region.
Using the current in Eq. (\ref{current}) and contracting the light quark fields in Eq. (\ref{masss}), we obtain the QCD side of the correlation function as
\begin{equation}
\label{corre1}
\Pi^{QCD}_{\mu\nu}= i \int d^4 x e^{ip\cdot x} Tr[\gamma_5 S^{ab}_u(x)\gamma_5\gamma_{\nu} S^{ba}_s(-x)\gamma_{\mu}],
\end{equation}
where  $S_{q=u(s)}$ is the full color carried light quark propagator in the dense medium. It can be expanded  as,
\begin{eqnarray}
\label{prop}
S^{kl}_q(x) &=& \Big(\frac{i}{2\pi^2}\frac{\not{x}}{ x^4}-\frac{m_q}{4\pi^2x^2}+i\frac{m_q^{2} }{8\pi^2 } \frac{\not{x}}{x^2}\Big)\delta^{kl}+ \chi_q^k(x) \bar{\chi}_q^l(0)\nonumber\\
&-& i\frac{g_s}{32 \pi^2}\frac{\not{x}\sigma_{\mu \nu}+\sigma_{\mu \nu}\not{x}}{x^2}F^{\mu \nu}_A (0) t_A^{kl}+...,
\end{eqnarray}
where the term inside the parenthesis  is the perturbative part and the second and the third terms represent the non-perturbative contributions.
Here, $\chi^k_q$ and $\bar{\chi}^l_q$ are the Grassmann background quark fields, $F^{\mu\nu}_A$ is the  background gluon field and  $t^{kl}_A=\frac{\lambda ^{kl}_A}{2}$ with  $\lambda ^{kl}_A$ being the  Gell-Mann matrices.

To proceed in QCD side we shall put the full expressions of the two propagators in Eq. (\ref{corre1}), which leads to a result containing a perturbative part and a nonperturbative part including different  in-medium  quark, gluon and mixed condensates, which are functions of the density of the nuclear matter ($ \rho $). We do not present all the steps in calculations, but refer the reader to, for instance,  Ref. \cite{Azizi:2014yea} for details. After using all the in-medium operators presented in  Ref. \cite{Azizi:2014yea}, we need to transfer  the calculations to the momentum space.  For this,  we use
\begin{eqnarray}
\frac{1}{(x^2)^n} &=& \int \frac{d^Dt}{(2\pi)^D}e^{-i t \cdot x} i(-1)^{n+1}2^{D-2n}\pi^{D/2} \nonumber \\
&&\times \frac{\Gamma[D/2-n]}{\Gamma[n]}\Big[-\frac{1}{t^2}\Big]^{D/2-n},
\end{eqnarray}
in $ D  $ dimension. We also use, $x_{\mu}\rightarrow -i\frac{\partial}{\partial p_{\mu}}$ for those $x_{\mu}  $ that appear in nominators of different terms. The next step is to perform the Fourier integral over $ x $, as a result of which a Dirac Delta function including the  momentum of the particle $ p $ and $ t $, originated from the above relation, appears.  Using the resultant Dirac Delta, we perform the D-dimensional integral over $ t $. This leads us to some expressions in momentum space. We set $ D=4-2 \epsilon $ and perform dimensional regularization. Then, we apply the standard Borel  transformation in order to suppress the contributions of the higher states and continuum. Finally, the following replacement is used to apply the continuum subtraction procedure:
\begin{equation}
\left( M^{2}\right) ^{N}e^{-m^{2}/M^{2}} \to
\frac{1}{\Gamma (N)}\int_{m^{2}}^{s_0}dse^{-s/M^{2}}\left( s-m^{2}\right)
^{N-1},
\end{equation}%
for $N>0$, where $ s_0 $ is the continuum threshold to be fixed based on the standard prescriptions of the method later. 
As a result of the above procedures, the following relations for the QCD sides of the vacuum and in-medium cases are obtained:
\begin{eqnarray}
\label{corre7}
\mathbf{ \hat{B}}[\frac{p_{\mu}p_{\nu}}{p^2}\Pi^{QCD}_{\mu\nu}] &=&\mathbf{\Omega}(M^2,s_0) \mathbf{ I}, \nonumber \\
\mathbf{ \hat{B}}\big[p^*_{\mu}\Pi^{QCD}_{\mu\nu}\big] &=&\mathbf{\Omega}^*_1(M^2,s_0,\rho) p_{\nu} +\mathbf{ \Omega}^*_2(M^2,s_0,\rho) u_{\nu},\nonumber\\
\end{eqnarray}
where  $\mathbf{ \Omega}(M^2,s_0)$, $\mathbf{ \Omega}^*_{1}(M^2,s_0,\rho)$ and $\mathbf{ \Omega}^*_{2}(M^2,s_0,\rho)$  are the Borel transformed amplitudes of the corresponding structures.
The expressions of these functions are very lengthy and we do not present their explicit forms for simplicity.

The sum rules for the mass and decay constant  of pseudoscalar state in vacuum and these quantities together with the vector self-energy of the particle in medium are obtained by equating the coefficients of the same structures from both the physical and QCD sides for each case.
In vacuum we obtain the sum rule
\begin{equation}
\label{vac}
 f_{K}^{2}m_{K}^{4} e^{-m_{K}^{2} /M^2}=\mathbf{\Omega}(M^2,s_0),
\end{equation}
 which  leads to
\begin{eqnarray}
m_{K}^{2}&=& \frac{\mathbf{\Omega'}(M^2,s_0)}{\mathbf{\Omega}(M^2,s_0)}, \nonumber \\
 f_{K}^{2}&=&\frac{\mathbf{\Omega}(M^2,s_0)}{m_{K}^{4}}e^{m_{K}^{2} /M^2},
\end{eqnarray}
for the mass and decay constant of the $ K $ meson in vacuum, where $\mathbf{\Omega'}(M^2,s_0)=\frac{d\mathbf{\Omega}(M^2,s_0)}{d(-1/M^2)} $.
We also obtain the following  density dependent sum rules for the mass, decay constant and vector self-energy of kaon in medium:
\begin{eqnarray}
 f_{PS}^{*2}m_{PS}^{*4} e^{-\mu^2/M^2}&=&\mathbf{\Omega}^*_1(M^2,s_0,\rho), \nonumber \\
 - f_{PS}^{*2}m_{PS}^{*4} \Sigma_{\upsilon}e^{-\mu^2/M^2}&=&\mathbf{\Omega}^*_2(M^2,s_0,\rho), \nonumber \\
   \Sigma_{\upsilon} &=& - \frac{\mathbf{\Omega}^*_2(M^2,s_0,\rho)}{\mathbf{\Omega}^*_1(M^2,s_0,\rho) }.
\end{eqnarray}
These sum rules are used to obtain the numerical values of the vacuum and in-medium physical quantities of kaon. For this, besides the mass of light quarks taken from Ref. \cite{PDG}, the parameters of vacuum and in-medium quark, gluon and mixed condensates with different dimensions and saturation nuclear matter density are needed (see for instance Ref. \cite{Azizi:2014yea}  and references therein) :
\begin{eqnarray}
&& m_u=2.16^{+0.49}_{-0.26} ~\textrm{MeV}, m_s=93^{+11}_{-5} ~ \textrm{MeV}, \nonumber \\
&& \rho^{sat}=(0.11)^3 \textrm{GeV}^3,  \langle q^{\dag} q\rangle_{\rho}=\frac{3}{2}\rho_N,  \langle s^{\dag} s\rangle_{\rho}=0, \nonumber \\
&&  \langle \bar{q}q\rangle_{0}=(-0.241)^3  \textrm{GeV}^3,  \langle \bar{s}s\rangle_{0} = 0.8 \langle \bar{q}q\rangle_{0}, \nonumber \\
&& \langle \bar{q}q\rangle_{\rho}=\langle \bar{q}q\rangle_{0}+\frac{\sigma_N}{2 m_q}\rho, \nonumber \\
&& \sigma_N=0.059~ \textrm{GeV}, m_q=0.00345 ~\textrm{GeV},  \nonumber \\
&& \langle \bar{s}s\rangle_{\rho}=\langle \bar{s}s\rangle_{0}+y\frac{\sigma_N}{2 m_q}\rho, y=0.05\pm0.01, \nonumber \\
&& \langle \frac{\alpha_s}{\pi}G^2\rangle_{0}=(0.33\pm0.04)^4 ~\textrm{GeV}^4,  \nonumber \\
&& \langle \frac{\alpha_s}{\pi}G^2\rangle_{\rho}=\langle \frac{\alpha_s}{\pi}G^2\rangle_{0}-(0.65\pm 0.15) ~\textrm{GeV} \rho, \nonumber \\
&& \langle q^{\dag}iD_0 q\rangle_{\rho}=0.18~\textrm{GeV}~\rho_N,  \nonumber \\
&& \langle s^{\dag}iD_0 s\rangle_{\rho}=\frac{m_s\langle \bar{s}s\rangle_{\rho}}{4}+0.02 ~\textrm{GeV} \rho, \nonumber \\
&&  \langle \bar{q}iD_0q\rangle_{\rho}=\langle \bar{s}iD_0s\rangle_{\rho}=0,  \nonumber \\
&& \langle \bar{q}g_s\sigma G q\rangle_{0}=m_0^2 \langle \bar{q}q\rangle_{0},  \langle \bar{s}g_s\sigma G s\rangle_{0}=m_0^2 ~\langle \bar{s}s\rangle_{0},   \nonumber \\
&& m_0^2=0.8  ~\textrm{GeV}^2, \langle \bar{q}g_s\sigma G q\rangle_{\rho}=\langle \bar{q}g_s\sigma G q\rangle_{0}+ 3~\textrm{GeV}^2~\rho,  \nonumber \\
&& \langle \bar{s}g_s\sigma G s\rangle_{\rho}=\langle \bar{s}g_s\sigma G s\rangle_{0}+ 3y~\textrm{GeV}^2~\rho,  \nonumber \\
&& \langle q^{\dag}g_s\sigma G q\rangle_{\rho}=-0.33 ~\textrm{GeV}^2~\rho,  \nonumber \\
&& \langle q^{\dag}iD_0 iD_0 q\rangle_{\rho}=0.031~\textrm{GeV}^2~\rho_N-\frac{1}{12}\langle q^{\dag}g_s\sigma G q\rangle_{\rho},  \nonumber \\
&&  \langle s^{\dag}g_s\sigma G s\rangle_{\rho}=-0.33y ~\textrm{GeV}^2~\rho,  \nonumber \\
&& \langle s^{\dag}iD_0 iD_0 s\rangle_{\rho}=0.031y ~\textrm{GeV}^2~\rho-\frac{1}{12}\langle s^{\dag}g_s\sigma G s\rangle_{\rho},
\end{eqnarray}
where $\langle  \mathcal{O} \rangle_0$, for the corresponding operator $ \mathcal{O}$,   is used  to represent the  vacuum condensates, while  $\langle  \mathcal{O} \rangle_{\rho}$ stands for the  in-medium condensates. 

\begin{figure}[h!]
\label{fig1a}
\centering
\begin{tabular}{c}
\epsfig{file=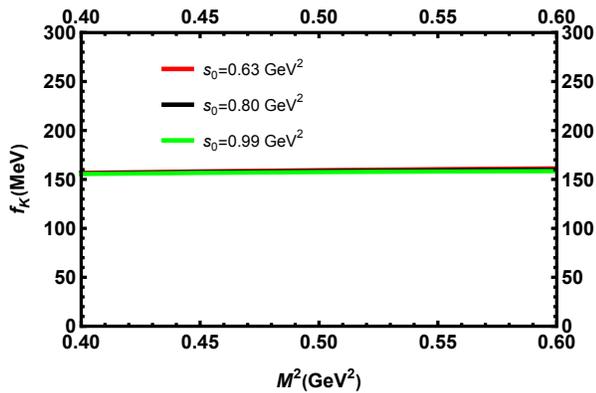,width=0.90\linewidth,clip=}
\end{tabular}
\caption{The decay constant of $ K^+ $ meson in vacuum with respect to Borel mass parameter at different continuum thresholds.}\label{Fig1}
\end{figure}

Some necessary conditions like pole dominance and convergence of the operator product expansion have to be fulfilled in order to fix the working regions for the auxiliary parameters $ M^2 $ and $ s_0 $. We also require that the dependence of the physical quantities on these auxiliary parameters be relatively weak. 
 As a result, we can fix the Borel and  continuum threshold parameters within the limits:
\begin{equation}
M^2 \in [0.4-0.6]~\textrm{GeV}^2 ~ \& ~ s_0 \in [0.63-0.99] ~\textrm{GeV}^2.
\end{equation}
Considering these intervals, we plot the $ M^2 $ dependence of the kaon decay constant in vacuum in Fig. (\ref{Fig1}). We observe that the decay constant shows a good stability against the variations of the auxiliary parameters within their working windows. Extracted from the analyses we depict the values of the mass and decay constant of the $ K^+ $ meson compared with the experimental data and Lattice QCD predictions  in Table \ref{tabbir}. We see  good consistencies among the presented values within the uncertainties.
\begin{table}[t]
{\begin{tabular}{lccc}\hline \hline
&  Method &   $m_{K}$ &  $f_{K}$  \\
\hline\hline
PS &  QCDSR  & $496.5^{+9.7}_{-10.8}$ MeV  & $158.1^{+3.2}_{-2.6}$ MeV \\
\cite{Carrasco:2014poa} & Lattice & - & $154.4(2.0)$ MeV \\
PDG \cite{PDG} & Exp & $493.677 \pm 0.016$ MeV & $155.6 (0.4)$ MeV \\
\cite{Ting} & Lattice & $478\pm 16\pm 20$ MeV  & $152\pm 6\pm 10$ MeV \\
\cite{Dimopoulos:2021qsf}  & Lattice & - & $156.3(0.6)$ MeV \\
\cite{flag} & Lattice & - & $155.7(0.3)$ MeV \\
\hline\hline
\end{tabular}}
\caption{Mass and decay constant of $ K^+ $ meson in vacuum compared with the Experimental data  and Lattice QCD  results. PS represents the presents study.}\label{tabbir}
\end{table}

\begin{figure}[h!]
\label{fig1}
\centering
\begin{tabular}{c}
\epsfig{file=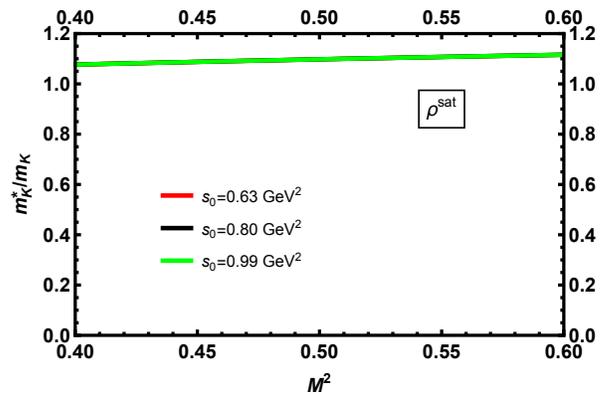,width=0.90\linewidth,clip=}
\end{tabular}
\caption{Variation of the ratio of the  in-medium  to vacuum mass for $ K^+ $ meson  with respect to Borel mass parameter at saturation density and different continuum thresholds.}\label{Fig2}
\end{figure}
\begin{figure}[h!]
\label{fig1}
\centering
\begin{tabular}{c}
\epsfig{file=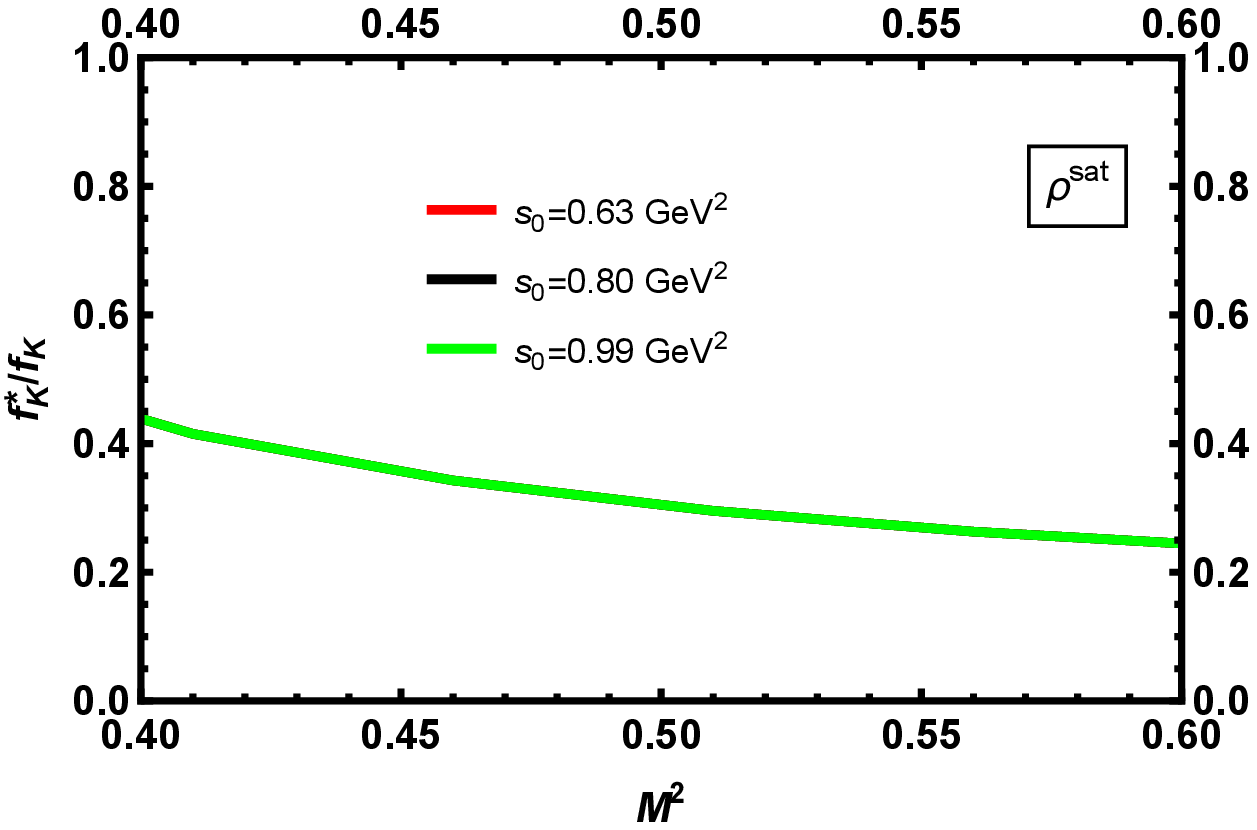,width=0.90\linewidth,clip=}
\end{tabular}
\caption{The same as Fig. \ref{Fig2} but for decay constant of $ K^+ $ meson.}\label{Fig3}
\end{figure}
\begin{figure}[h!]
\label{fig1}
\centering
\begin{tabular}{c}
\epsfig{file=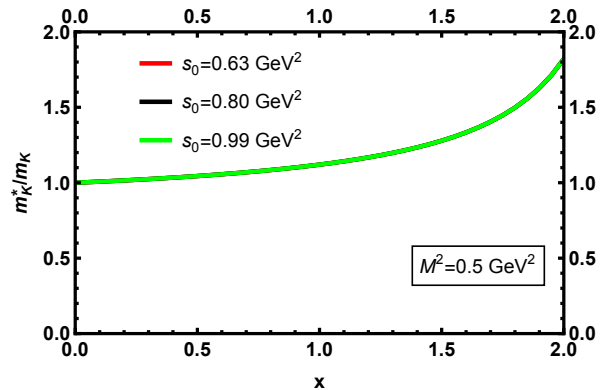,width=0.90\linewidth,clip=}
\end{tabular}
\caption{The density dependence of the ratio of in-medium to vacuum mass  for $ K^+ $ meson at  $M^2=0.5$ GeV$^2$  and different continuum thresholds.}\label{Fig4}
\end{figure}
\begin{figure}[h!]
\label{fig1}
\centering
\begin{tabular}{c}
\epsfig{file=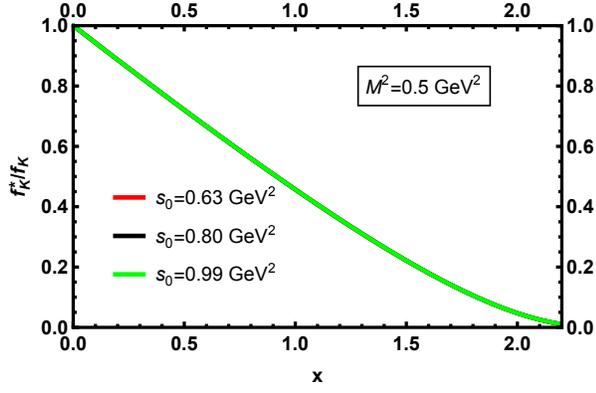,width=0.90\linewidth,clip=}
\end{tabular}
\caption{The same as Fig. \ref{Fig4} but for decay constant of $ K^+ $ meson.}\label{Fig5}
\end{figure}

We proceed to discuss the dependencies of the considered quantities on the density of the nuclear medium. To check the stability of the results with respect to the auxiliary parameters,  in Figs. (\ref{Fig2}) and (\ref{Fig3}), we depict the rations $m^*_K/m_K $ and $ f^*_K/f_K$   as functions of $M^2$ at different values of  $s_0$ at the saturation nuclear matter density $\rho^{sat}$. From these figures, we  see a good stability of $m^*_K/m_K $ and $ f^*_K/f_K$   with respect to the continuum threshold. The ratio $m^*_K/m_K $ shows a nice stability with respect to $M^2$, as well. We see some residual dependence of $ f^*_K/f_K$ to $M^2$ but it remains within the limits allowed by the QCD sum rule method. 

The last step in this section is to discuss the dependence of the physical quantities under consideration on the density of the nuclear medium. To this end, in Figs. (\ref{Fig4}) and (\ref{Fig5}) we plot the dependence of the rations $m^*_K/m_K $ and $ f^*_K/f_K$    on $ x=\rho/\rho^{sat} $. These figure indicate that the mass of $ K^+ $ meson increases with the increase in the density, although the rate at lower densities is small. The decay constant linearly decreases with the increase in the density of the medium and it reaches to zero at $ \rho\simeq 2.2 \rho^{sat} $  and we are witness of the melting of meson at higher densities. At saturation density, the result of the present study for  the in-medium mass is compared with the result of Chiral SU(3) model  obtained in Ref. \cite{Mishra:2018ogw} (see Table \ref{tab2}): Both show  considerable growths with respect to the medium mass value. We shall use the density-dependent functions of the mass and decay constant as main inputs in next section to determine the density-dependent electromagnetic form factor and other observables.

\begin{table}[t]

{\begin{tabular}{lcc}\hline \hline \\
&  Method &  $m^{*}_{K}$  \\
\hline\hline
PS &  QCDSR    & $553\pm12$ MeV  \\
\cite{Mishra:2018ogw} & Chiral SU(3) model   & $521$ MeV \\
\hline\hline
\end{tabular}}
\caption{The in-medium mass of kaon at saturation density.}\label{tab2}
\end{table}

\section{Electromagnetic form factor and charge radius of Kaon in vacuum and a dense medium}

The following in-medium three-point correlation function is responsible  for extraction of the  electromagnetic form factor and charge radius:
\begin{eqnarray}
\label{elec}
\Pi_{\mu\nu\lambda} &=& i^2 \int d^4 x  \int d^4 y e^{ip\cdot x} e^{-ip'\cdot y}\nonumber\\
&\times&\langle\psi_0|\mathcal{T}[j_{\mu}(y)J_{\lambda}(0)j{\dagger}_{\nu}(x)]|\psi_0\rangle,
\end{eqnarray}
where $ p=p'+q $ and $ p' $ are the momenta of the initial and final kaonic states, respectively and $ q $ is carried by the electromagnetic current.   The electromagnetic current is
\begin{equation}
J_{\lambda}(0)= e_u \bar{u}(0)\gamma_{\lambda}u(0) + e_d \bar{d}(0) \gamma_{\lambda}d(0)  + e_s \bar{s}(0) \gamma_{\lambda}s(0), 
\end{equation}
with $e_{u,d,s}$ being the charge of the corresponding quark. Using the axial current and current dagger at two different points for kaon  and the electromagnetic current we get the following expression in terms of the quark propagators after applying the Wick's theorem:
\begin{eqnarray}
\label{corre2}
&&\Pi^{QCD}_{\mu\nu\lambda} =i^2  \int d^4 x  \int d^4 y e^{ip\cdot x} e^{-ip'\cdot y}\nonumber\\ &\times &\Big[e_u Tr[S^{da}_s(x-y)\gamma_{\mu}\gamma_5 S^{ab}_u(y)\gamma_{\lambda} S^{bd}_u(-x)\gamma_5 \gamma_{\nu}]\nonumber \\
&+&e_s[S^{dc}_s(x)\gamma_{\lambda} S^{ca}_s(-y)\gamma_{\mu} \gamma_5 S^{ad}_u(y-x)\gamma_5 \gamma_{\nu}]\Big]_{\psi_0}.\nonumber\\
\end{eqnarray}
We apply similar procedures as the previous section: The full light quark propagators in medium and vacuum are placed in the above equation and the standard Fourier, double-Borel and continuum subtraction procedures are performed. For simplicity we do not present these lengthy but standard procedures here. As a result of the above procedures, the correlation function in QCD side is obtained. 

The physical side of the calculations is obtained by saturation of the correlation function (\ref{elec}) with two complete sets of intermediate states both in the $ AV  $ and $ PS $ channels. Performing the integrals over four-$ x $ and four-$ y $, we get
\begin{widetext}
\begin{eqnarray}
\label{corre4}
\Pi^{Phys}_{\mu\nu\lambda} =&&\frac{ \langle\psi_0|j_{\mu}|K^{AV}(p^*_2)\rangle  \langle K^{AV}(p^*_2) | J_{\lambda}| K^{AV}(p^*_1) \rangle \langle K^{AV}(p^*_1) | j^{\dagger}_{\nu}|\psi_0\rangle}{(m_{AV}^{*2}-p_1^{*2})(m_{AV}^{*2}-p_2^{*2})} \nonumber \\
&+&\frac{ \langle\psi_0|j_{\mu}|K^{PS}(p^*_2)\rangle  \langle K^{PS}(p^*_2) | J_{\lambda}| K^{PS}(p^*_1) \rangle \langle K^{PS}(p^*_1) | j^{\dagger}_{\nu}|\psi_0\rangle}{(m_{PS}^{*2}-p_1^{*2})(m_{PS}^{*2}-p_2^{*2})}+ ...,
\end{eqnarray}
in medium, where $ p^*_1=p^* $ and $ p^*_2=p ^{'*}$. The vacuum version of this step  is obtained by replacing the star quantities with their without star versions. The transition martix elements in Eq. (\ref{corre4}) are defined in terms of the in-medium form factors as
\begin{eqnarray}
\label{form}
 \langle K^{AV}(p^*_2) | J_{\lambda}| K^{AV}(p^*_1) \rangle &=& \varepsilon'_{\tau}(p^*_2)\Bigg[F_1^{*AV}(Q^{2}) g_{\tau\sigma}\big(p_{1,\lambda}^*, +p_{2,\lambda}^*\big) +F_2^{*AV}(Q^{2})\big(q^*_{\tau} g_{\lambda\sigma}-q^*_{\sigma} g_{\lambda\tau}\big) \nonumber \\
 &-&\frac{F_3^{*AV}(Q^{2}) q^*_{\tau}q^*_{\sigma}\big(p_{1,\lambda}^*+p_{2,\lambda}^*\big)}{2 m_{AV}^{*2}}\Bigg]\varepsilon_{\sigma}(p^*_1),\nonumber \\
 \langle K^{PS}(p^*_2) | J_{\lambda}| K^{PS}(p^*_1) \rangle &=&F^{*PS}(Q^{2})\big(p_{1,\lambda}^*+p_{2,\lambda}^*\big),
\end{eqnarray}
where $Q^{2}=-q^{2}  $; and $\varepsilon$ and $\varepsilon'$ are the polarization vectors of the initial and final axial states, respectively. Here, $F_{1,2,3}^{*AV}(Q^{2})$ are the axial vector particle and $F^{*PS}(Q^{2})$ is the pseudoscalar particle invariant electromagnetic form factors. By using Eq. (\ref{form}) in Eq. (\ref{corre4}), one obtains an expression for $  \Pi^{Phys}_{\mu\nu\lambda}$ in terms of the in-medium form factors,  polarization vectors, momenta and other hadronic parameters. To remove the axial vector particle contribution we multiply  both the  physical and QCD sides of the correlation function by  $\frac{p^*_{1\mu}p^*_{1\nu}}{p_1^{*2}}$. The next step is to  put the in-medium momenta in terms of the external momenta, vector self-energy and four-velocity vector of the medium like the previous section. We then  apply the double Borel transformation with respect to $ p_1^2 $ and  $ p_2^2 $ and use
$ \mu_1^2=  \mu_2^2= \mu^2$, $ \frac{1}{M^2} =\frac{1}{M_1^2}+\frac{1}{M_2^2}$ and $ M_1^2=M_2^2=2M^2 $ (the initial and final particles are the same). As a result, we get 
\begin{equation}
\label{effcor}
\mathbf{ \hat{B}}\Big[\frac{p^*_{1\mu}p^*_{1\nu}}{p_1^{*2}}\Pi^{Phys}_{\mu\nu\lambda}\Big] = f_{PS}^{*2}F^{*PS}(Q^2)(2m_{PS}^{*2}+Q^2) e^{-\mu^2/M^2} \Big(\frac{p_{1\lambda}+p_{2\lambda}-2\Sigma_{\upsilon} u_{\lambda}}{2}\Big),
\end{equation}
\end{widetext}
which contains three $p_{1\lambda}  $,  $p_{2\lambda}  $ and $u_{\lambda}  $ structures. We use the structure  $p_{1\lambda}  $ to extract the electromagnetic form factor of the pseudoscalar $ K^+ $ particle. Similarly, we  can rewrite the final expression of the QCD side  in terms of the same Lorentz structures in Borel scheme as:
\begin{eqnarray}
\label{corre7}
&&\mathbf{ \hat{B}}[\frac{p^*_{1\mu}p^*_{1\nu}}{p_1^{*2}}\Pi^{QCD}_{\mu\nu\lambda}(Q^2)] =\mathbf{\Theta}_1(M^2,s_0,\rho) p_{1\lambda}  \nonumber\\&&+\mathbf{\Theta}_2(M^2,s_0,\rho) p_{2\lambda}+\mathbf{\Theta}_3(M^2,s_0,\rho) u_{\lambda},~
\end{eqnarray}
where $\mathbf{\Theta}_{i=1,2,3}(M^2,s_0,\rho)$ are the Borel transformed amplitudes, whose expressions are very lengthy and we prefer to not present their explicit forms here. 

By equating the coefficients of the structure $p_{1\lambda}  $, from both  the physical and QCD sides, the sum rule for the in-medium form factor $ F^{*PS}(Q^2) $ is obtained in terms of the the in-medium mass and decay constant, the auxiliary parameters $ M^2 $ and $ s_0 $ as well as other QCD degrees of freedom like the quark masses and quark-gluon in-medium condensates. The vacuum form factor $ F^{PS}(Q^2) $ is obtained from $ F^{*PS}(Q^2) $ in the limit $ \rho\rightarrow 0 $.

\begin{figure}[h!]
\label{fig1}
\centering
\begin{tabular}{c}
\epsfig{file=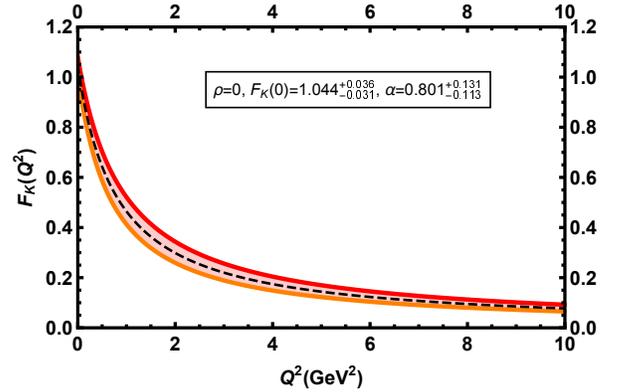,width=0.90\linewidth,clip=}
\end{tabular}
\caption{The vacuum electromagnetic form factor of Kaon in terms of  $ Q^2 $ considering all the uncertainties of the input parameters.}\label{Fig6}
\end{figure}

 Let us first discuss the $ Q^2 $ dependence of the vacuum form factor $ F^{PS}(Q^2) $ in the interval $ Q^2\in[0,10] $ GeV$^2$ and compare the obtained results with the existing experimental data as well as other theoretical predictions. Using all the input parameters we find that   the following  monopole function  explains the  form factor $ F^{PS}(Q^2) $ (hereafter we omit the super-index $ PS $ and add the sub-index $ K $) in terms of $ Q^2 $:
\begin{equation}
F_K(Q^2)=\dfrac{F_K(0)}{\Big[1+\dfrac{Q^2}{\alpha}\Big]},
\end{equation}
where  $F_K(0)=1.044^{+0.036}_{-0.031}$ and $\alpha=0.801^{+0.131}_{-0.113}$ GeV$^2$ are obtained. We plot the vacuum form factor $ F_{K}(Q^2) $  in terms of $ Q^2 $ in Fig. (\ref{Fig6}), where the the uncertainties are appeared as a band. We see that the form factors falls with increasing the value of $ Q^2 $ and reaches to roughly  $ 10\% $ of  $F_K(0)  $ value at $ Q^2 =10$ GeV$^2$. 
\begin{figure}[h!]
\label{fig1}
\centering
\begin{tabular}{c}
\epsfig{file=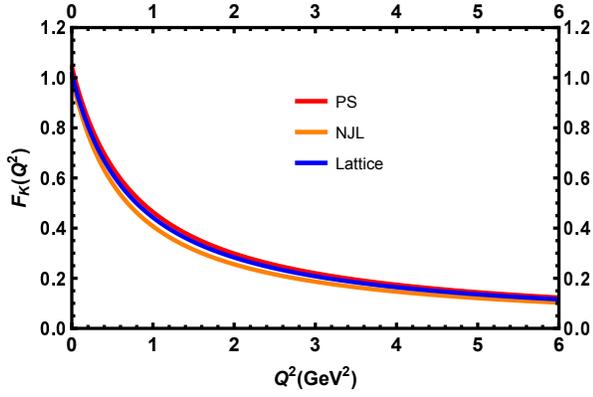,width=0.90\linewidth,clip=}
\end{tabular}
\caption{Comparison of the kaon electromagnetic form factor with different studies (orange solid line \cite{Hutauruk} and blue solid line \cite{Alexandrou:2021ztx}).}\label{Fig7}
\end{figure}
\begin{figure}[h!]
\label{FQsq}
\centering
\begin{tabular}{c}
\epsfig{file=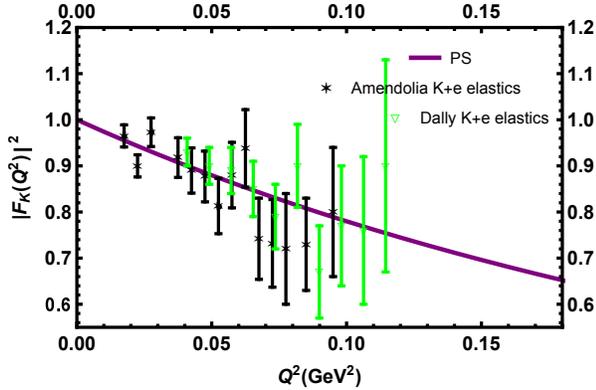,width=0.90\linewidth,clip=}
\end{tabular}
\caption{Variation of $ \vert F_{K}(Q^2\vert^2 $ at low $Q^2$ compared to existing experimental data \cite{Amendolia:1986ui,Dally:1980dj} with $|F_K(Q^2=0)|^2$ normalized to one.}\label{Fig8}
\end{figure}
\begin{figure}[h!]
\label{FFQssq}
\centering
\begin{tabular}{c}
\epsfig{file=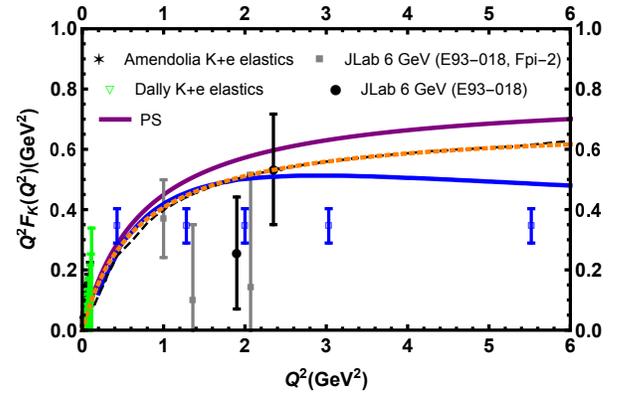,width=0.90\linewidth,clip=}
\end{tabular}
\caption{$ Q^2 F_{K}(Q^2)$  with respect to  $Q^2$. The result of  present study (PS) is represented with the solid purple line. The black stars and green triangles show experimental data taken from Refs. \cite{Amendolia:1986ui,Dally:1980dj}. The filled squares corresponds to kaon electromagnetic form factor from E93-018 and FPI-2. The filled circles data were extracted form the cross section data of Refs. \cite{Coman,Horn:2016rip}. Blue points with error bars are not data points, they represent the $Q^2$ range and projected uncertainties of the E12-09-011 JLab experiment \cite{Horn:2016rip}. The dashed black curve shows the predictions of Refs. \cite{Amendolia:1986ui,PhysRevC.97.025204}. The dotted orange line is borrowed from Ref. \cite{Hutauruk}. The solid blue line shows the kaon form factor obtained in Ref. \cite{PhysRevD.96.034024}.}\label{Fig9}
\end{figure}
In Fig. (\ref{Fig7}), we compare our result on $ F_{K}(Q^2)- Q^2$ with the existing prediction from a model based on the solution of the Bethe-Salpeter equation for the model of NJL with proper-time regularization  \cite{Hutauruk} and Lattice QCD  \cite{Alexandrou:2021ztx} in the interval $ Q^2\in[0,6] $ GeV$^2$. From this figure, we see a nice consistency among our result and the other predictions. In Fig. (\ref{Fig8}), we compare our result on $ \vert F_{K}(Q^2\vert^2 $ with respect to $Q^2$ at lower values of $Q^2$. Within the presented uncertainties, we again see a nice agreement between our result and the presented two sets of the experimental data. Finally, in Fig. (\ref{Fig9}), we compare our result on the behavior of 
$ Q^2 F_{K}(Q^2)$  with respect to  $Q^2$ in the interval $ Q^2\in[0,6] $ GeV$^2$ with the existing data and predictions of other theoretical and phenomenological models.  Although we see a good consistency among all the presented results for $ Q^2\leq 1 $ GeV$^2$, for $ Q^2 >1 $ GeV$^2$
we observe overall considerable discrepancies among the theoretical predictions and experimental data that requires more theoretical and experimental studies on the considered quantity at mean values of $Q^2$. Our results show relatively a good consistency with the results of  Refs. \cite{Amendolia:1986ui,PhysRevC.97.025204,Hutauruk} at middle values of $Q^2$, as well.

The final subject related to the vacuum analyses is the charge radius. The electromagnetic charge radius of the K meson is obtained from the slope of the electromagnetic form factor at $Q^2=0$ value, i.e., 
\begin{equation}
\langle r_K^2 \rangle = -6 \dfrac{d}{dQ^2}F_K(Q^2)\Big|_{Q^2=0}.
\end{equation}
Table \ref{tab3} compares our result for charge radius in vacuum  with the experimental data (world average) and other phenomenological predictions. From this table, we observe that our prediction for  $r_{K}$ is in a nice agreement with the world average experimental data. It is also consistent with the predictions of Refs. \cite{Hutauruk,Ebert}. The remaining predictions are sightly higher than our result and the experimental data. 

\begin{figure}[h!]
\label{fig1}
\centering
\begin{tabular}{c}
\epsfig{file=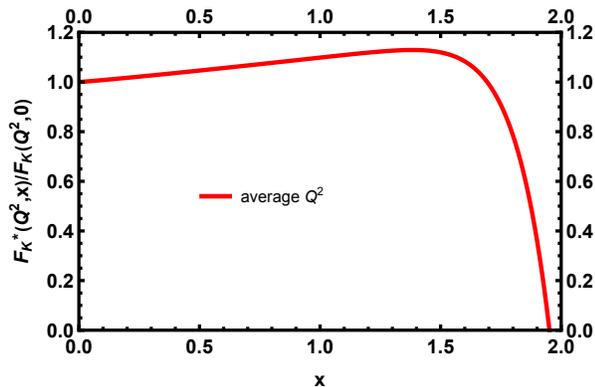,width=0.90\linewidth,clip=}
\end{tabular}
\caption{The variation of the ratio of the density dependent electromagnetic form factor to the vacuum one with respect to the density of the medium at average value of $ Q^2 $.}\label{Fig10}
\end{figure}

\begin{figure}[h!]
\label{fig1}
\centering
\begin{tabular}{c}
\epsfig{file=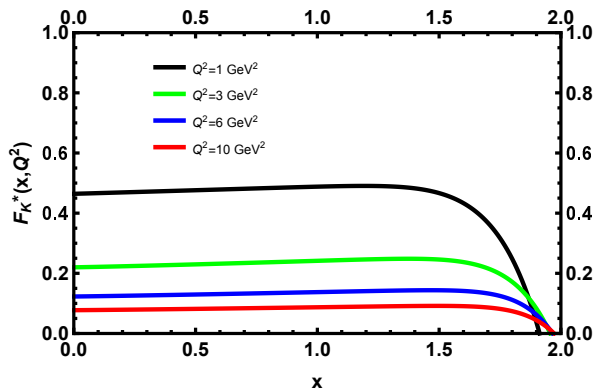,width=0.90\linewidth,clip=}
\end{tabular}
\caption{The variation of  $F_K^*(Q^2,x)$  with respect to the density of the medium at different values of $ Q^2 $.}\label{Fig11}
\end{figure}

Now, we proceed to discuss the dependence of the electromagnetic form factor and charge radius of the $ K^+ $ meson on the density.  For this aim, in Fig. (\ref{Fig10})  we plot the ratio $\frac{F_K^*(Q^2,x)}{F_K(Q^2,0)}$ as a function of $ x=\rho/\rho^{sat} $ at average $ Q^2 $. In Fig. (\ref{Fig11}), we depict the  dependence of $  F_K^*(Q^2,x)$ on $ x $ at different values of $ Q^2 $ and at average values of the auxiliary parameters. From these figures, we observe that the in-medium form factor of kaon starts to increase slightly, however, after $ \rho=1.5  \rho^{sat} $ it starts to diminish and sharply goes to zero at a density about $2  \rho^{sat} $. 
\begin{table}[t]
{\begin{tabular}{lcc}\hline \hline
&  Method &  $r_{K}$  \\
\hline\hline
PS &  QCDSR    & $0.54\pm 0.05$ fm  \\
\cite{PhysRevC.86.038202} & LFCQM & $0.636$ fm \\
\cite{Maris:2000sk}  & DSEs & $\sqrt{0.38}$ fm \\
\cite{Krutov:2016luz}& Fit to NA-7 data & $\sqrt{0.39} \leq r_K \leq  \sqrt{0.42}$ fm \\
PDG \cite{PDG} & Exp   & $0.560\pm 0.031$ fm \\
\cite{Hutauruk} & NJL  & $0.586$ fm \\
\cite{Aoki}& Lattice & $\sqrt{0.380(12)(^{+7}_{-1})(31)}$ fm \\
\cite{Bijnens:2002hp}& CPT & $ \sqrt{0.431\pm 0.071}$ fm \\
\cite{Ebert} & RQM & $0.57$ fm \\
\hline\hline
\end{tabular}}
\caption{Experimental result and theoretical predictions of different methods for the kaon  vacuum charge radius.}\label{tab3}
\end{table}
\begin{figure}[h!]
\label{fig1}
\centering
\begin{tabular}{c}
\epsfig{file=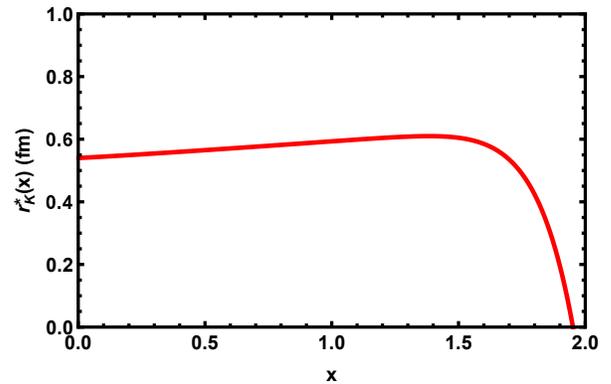,width=0.90\linewidth,clip=}
\end{tabular}
\caption{Variation of the charge radius of the kaon with respect to the density of the medium.}\label{Fig12}
\end{figure}
\begin{figure}[h!]
\label{fig1}
\centering
\begin{tabular}{c}
\epsfig{file=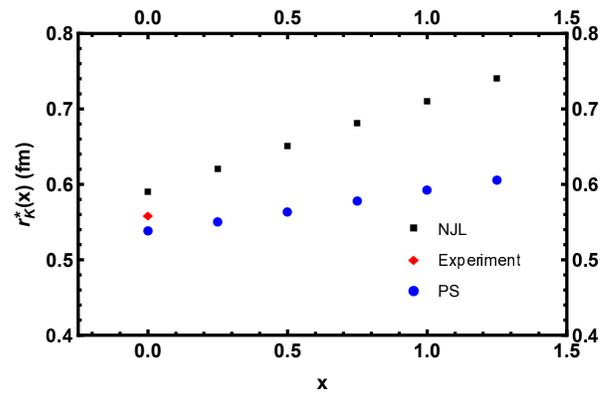,width=0.90\linewidth,clip=}
\end{tabular}
\caption{Comparison of our results for kaon charge  radius (blue circle points) with predictions of  Ref. \cite{Hutauruk:2019was} using NJL model (black squares) at different densities. We also show the experimental result in vacuum.  }\label{Fig13}
\end{figure}

In Fig. (\ref{Fig12}) we present the variation of the charge radius of the kaon with respect to the density of the medium. From this figure, we observe that, like the in-medium form factor, the radius is slightly increased with increasing in the density, but, it starts to fall after  $ \rho=1.5  \rho^{sat} $ point and sharply goes to zero at about the density $2  \rho^{sat} $. Finally in Fig. (\ref{Fig13}), we compare our results for the in-medium  kaon charge  radius, $r_K^*(x)$, with existing predictions of  Ref. \cite{Hutauruk:2019was} using NJL model  at different densities. The experimental data (red diamond)  at zero density is also added for a better comparison. From this figure we see that the behavior of charge radius with respect to the density is similar in  our study and Ref. \cite{Hutauruk:2019was} in the interval $ x<1.5 $: both predictions growth with increasing in the density, although our data points growth gradually compared to the data points obtained in \cite{Hutauruk:2019was} which show larger slope. As we previously mentioned, our prediction for the charge radius in vacuum is in a nice agreement with the experimental value in vacuum. 

\section{Summary and concluding notes}
The spectroscopic parameters as well as the electromagnetic form factor and charge radius of $ K^+ $ meson were studied in the context of QCD sum rules both in the vacuum and a medium with finite density. In vacuum, we obtained  the mass and decay constant of $ K^+ $ meson in a nice consistency with the experimental result and  Lattice QCD predictions. 

The electromagnetic form factor of this particle was found to have a form of monopole with respect to  $ Q^2 $. We investigated 
the behavior of $F_K(Q^2)$ in terms of $ Q^2 $ in a wide range, namely in the interval $ Q^2\in[0,10] $ GeV$^2$. We compared our result with the existing experimental data and other theoretical prediction. We saw  a nice consistency among our result and the model based on the solution of the Bethe-Salpeter equation for   NJL model with proper-time regularization  \cite{Hutauruk} as well as  Lattice QCD  \cite{Alexandrou:2021ztx} in the interval $ Q^2\in[0,6] $ GeV$^2$. Within the presented uncertainties, we also observed  a nice agreement among our result and the  two sets of the experimental data, namely Amendolia and Dally at low $ Q^2 $s. We also compared our result on $ Q^2 F_{K}(Q^2)$  with respect to  $Q^2$ in the interval $ Q^2\in[0,6] $ GeV$^2$ with the existing data and predictions of other theoretical and phenomenological models.  Although we observed a good consistency among all the presented results for $ Q^2\leq 1 $ GeV$^2$, for $ Q^2 >1 $ GeV$^2$ we saw overall considerable discrepancies among the theoretical predictions and experimental data that requires more theoretical and experimental studies. Our results show relatively a good consistency with the results of  Refs. \cite{Amendolia:1986ui,PhysRevC.97.025204,Hutauruk} at middle  values of $Q^2$, as well.

We also calculated the charge radius $r_{K}$ in vacuum and compared the obtained result with the experimental data  and other phenomenological predictions.We observed that our prediction for  $r_{K}$ is in a nice agreement with the world average experimental data. It is also consistent with the predictions of Refs. \cite{Hutauruk,Ebert}. 

Regarding the dense medium, we  discussed  the dependence of the mass and decay constant of $ K^+ $ meson on the density of the nuclear medium. To this end, we plotted the dependence of the rations $m^*_K/m_K $ and $ f^*_K/f_K$    on $ x=\rho/\rho^{sat} $. We concluded that the mass of $ K^+ $ meson increases with the increase in the density, although the rate at lower densities is small. The decay constant linearly decreases with the increase in the density of the medium and it approaches  to zero at $ \rho\simeq 2.2 \rho^{sat} $. At saturation density, the result of the present study for  the in-medium mass was compared with the result of Chiral SU(3) model  obtained in Ref. \cite{Mishra:2018ogw}: Both show  considerable growths with respect to the medium mass value. Our results on the behavior of the in-medium mass and decay constant may help experimental groups aiming to study the in-medium properties of hadrons. 

We also discussed the dependence of the electromagnetic form factor and charge radius of the $ K^+ $ meson on the density of the nuclear matter.  For this aim,  we plotted the ratio $\frac{F_K^*(Q^2,x)}{F_K(Q^2,0)}$ as a function of $ x=\rho/\rho^{sat} $ at average $ Q^2 $.  We also discussed the  dependence of $  F_K^*(Q^2,x)$ on $ x $ at different values of $ Q^2 $ and at average values of the auxiliary parameters. As a result,  we observed that the in-medium form factor of kaon starts to increase slightly, however, after $ \rho=1.5  \rho^{sat} $ it starts to diminish and sharply goes to zero at a density about $2  \rho^{sat} $.

Finally, we presented the variation of the charge radius $r_K^*$ with respect to the density of the medium. We observed that the radius slightly increases with increasing in the density, however, it starts to fall after  $ \rho=1.5  \rho^{sat} $ point and approaches  to zero at  $\rho=2  \rho^{sat} $. We also compared our results for $r_K^*$ at different densities with existing predictions of  NJL model \cite{Hutauruk:2019was} .  We concluded  that the behavior of $r_K^*$ with respect to the density is similar in  our study and the NJL model  and  both predictions growth with increasing in the density.  Once more we shall note that our prediction for the charge radius in vacuum is in a nice agreement with the world average experimental data. 

The results obtained in the present study, especially, those obtained at finite densities may be checked via different phenomenological models and approaches. They can  be useful in analyses of the future related data to be provided by the in-medium experiments. Comparison of the obtained vacuum and in-medium results in the present study with the existing and future experimental data can provide useful information about the nature, internal structure, size and inner charge distribution of the strange meson, $ K $, as well as the nonperturbative nature of QCD as  theory of the strong interaction.

\section*{ACKNOWLEDGEMENTS}
K. Azizi is thankful to Iran Science Elites Federation (Saramadan)
for the partial  financial support provided under the grant number ISEF/M/400150.



\end{document}